\documentclass{article}
\usepackage[utf8]{inputenc}
\usepackage[margin=2.5cm]{geometry}
\usepackage{authblk}
\usepackage{graphicx}
\usepackage{tabularx}
\usepackage{mhchem}
\usepackage{lineno}
\usepackage{siunitx}
\usepackage{xr}
\usepackage{hyperref}
\usepackage{cleveref}

\begin{document}

\title{Water Nucleation via Transient Bonds to Oxygen Functionalized Graphite}

\author[1,*]{Frederik G. Doktor}

\author[1]{Niels M. Mikkelsen}

\author[1]{Signe Kyrkjebø}

\author[3]{Prashant Srivastava}

\author[1]{Richard Balog}

\author[1]{Bjørk Hammer}

\author[3]{Karina Morgenstern}

\author[1,2]{Liv Hornek\ae r}

\affil[1]{Center for Interstellar Catalysis (InterCat), Department of Physics and Astronomy, Aarhus University, Ny Munkegade 120, 8000 Aarhus C, Denmark}
\affil[2]{Interdisciplinary Nanoscience Center (iNANO), Aarhus University, Gustav Wieds Vej 14, 8000 Aarhus C, Denmark}
\affil[3]{Faculty of Chemistry and Biochemistry, Physical Chemistry I, Ruhr
University of Bochum, Universitätsstraße 150, D-44801 Bochum, Germany}
\affil[*]{Frederik\_doktor@phys.au.dk}

\maketitle
\begin{abstract}
 We present a study the initial stages of ice growth on pristine and oxygen-functionalized highly oriented pyrolytic graphite (O-HOPG), combining low-temperature scanning tunneling microscopy (LT-STM) and machine-learning structural searches. LT-STM images show that oxygen atoms act as nucleation sites for ice growth, and that the size, structure and porosity of the nanometer-sized ice clusters depend strongly on the growth temperature. Machine learning-assisted structural searches and first-principles energy calculations confirm that clusters of water molecules are likely to bind to chemisorbed oxygen atoms through hydrogen bonding. During the early stages of the cluster growth clusters of water molecules are likely to be immobilized by binding to more than one chemisorbed oxygen atom through hydrogen bonding. However, the energy gain by hydrogen bond formation of a molecule, upon incorporation into smaller clusters only bound to a single oxygen atom, is large enough to induce cluster diffusion and favor the growth of larger ice clusters. Our results demonstrate that the mobility of water molecules is significantly lowered in the presence of defects on the surface. The observed lower mobility on defected carbon presented here offers an enhanced understanding of macroscopic anti-icing properties observed for functionalized HOPG under ambient conditions and provides insight into the early stages of ice growth on dust grain surfaces in interstellar space.

\end{abstract}

\section{Introduction}
The interaction of water with carbon surfaces is of interest in a wide range of fields such as energy storage, corrosion, anti-icing and biomedical engineering. A key goal across these fields is to control the wetting properties of typical sp²-carbon materials, including graphene, graphite, nanotubes, and fullerenes, to enable their use in electrodes, coatings, and membranes \cite{bae2012towards,kyhl2015graphene,gkika2023membranes}. Graphene and functionalized graphene have recently attracted interest as potential coating for anti-icing purposes \cite{Akhtar2019,Kyrkjebo2021}. Kyrkjebø \textit{et al}. showed that by adding oxygen to graphene-covered Ir(111), the freezing onset temperature of a water droplet was lowered from \qty[separate-uncertainty = true]{-15(1)}{\celsius} for bare graphene-Ir(111) to \qty[separate-uncertainty = true]{-21(1)}{\celsius} \cite{Kyrkjebo2021}. The lower freezing onset was suggested to originate from water being more viscous near the chemical defects \cite{Kyrkjebo2021,Zokaie2015}, which increases the barrier for a phase change according to ice nucleation theory \cite{Li2014}. However, the molecular interactions that drive this mechanism are not well understood.

Ice nucleation and growth on carbonaceous surfaces is not only relevant in terrestrial settings but is also an important topic of study in astrochemistry. In the cold regions of the interstellar medium (ISM) - the area of space in between the stars - ices form, nucleate, and grow on nanoscale silicate and carbonaceous dust grains \cite{willner1982infrared,herbst2014three}. Chemical reactions driven by radicals and energetic processing take place in and on the surface of these ices, as well as at the interface between the ice and the dust grain surface, resulting in the formation of complex organic molecules (COMs) with potential prebiotic significance \cite{Fulvio2021, Ioppolo2021}. The dynamics of water clusters and how they may form and stabilize on graphitic surfaces under interstellar conditions is a key step in understanding the early build up of ice on dust grains in the ISM and their role as catalysts for the emergence of molecular complexity in these regions \cite{burke2010ice,cazaux2016dust}. In addition, for highly porous dust grains it has been proposed that even in dense interstellar clouds ice layers will only be a few monolayers thick \cite{Potapov2020}. As a consequence, reactions at the dust-ice interphase may contribute significantly to interstellar chemistry \cite{potapov2019evidence}. Carbonaceous interstellar dust grain surfaces are expected to be highly defected. The properties of carbon surfaces are altered greatly when defects are introduced affecting both binding energies and mobility of adsorbed water molecules. However, the effect that such defects have on low-temperature ice nucleation is yet to be understood at the molecular level. 

Heterogeneous surfaces, \textit{i.e.}, surfaces with varying site-dependent binding energies, have been observed to alter ice growth on the microscopic level. E.g., Standop \textit{et al}. have recently demonstrated, using low-temperature scanning tunneling microscopy (LT-STM), that ice nucleates at hydrophilic areas on the surface of graphene on Ir(111), where the distance between the graphene layer and Ir(111) is smaller \cite{Standop2015}. Similar effects have been predicted \cite{Cao2011} and observed \cite{Zhang2020} on highly-oriented pyrolytic graphite (HOPG), where atomic step-edges initiate ice growth. Hence, variations in surface properties, such as roughness and composition, influence ice nucleation and growth significantly. Microscopic surface characteristics might therefore impact both, the macroscopic icing properties of surfaces as well as the microscopic ice nucleation and growth on dust grains in the ISM, ultimately affecting their catalytic activity.

In this article, we investigate the kinetics of ice nucleation and growth on bare and oxydized HOPG (O-HOPG) at cryogenic temperatures below \qty{80}{\kelvin} via low-temperature scanning tunneling microscopy (LT-STM). The experimental results are compared to computational results of the structure and dynamics of small water clusters on graphite, identified through machine learning-accelerated structural searches with first-principles energy calculations.
\section{Methods}

\subsection*{Experimental methods}
The experiments were carried out at two separate setups allowing us to explore the range of temperatures, surfaces, and water fluxes presented in this article. Both ultra-high vacuum (UHV) setups feature a preparation chamber connected to a low-temperature scanning tunneling microscope (LT-STM) which is cooled via liquid helium (LHe) from an isolated bath cryostat, see sketches of the two chambers, components, and deposition techniques in the supporting information. One setup is capable of cooling its manipulator arm with LHe, ensuring a low surface temperature of \qty{\sim45}{K} during sample preparation and the LT-STM itself reaches \qty{\sim8}{K}. A molecule deposition chamber, connected to the preparation chamber via a gate valve, is used to directly deposit the water onto the surface of interest. Attached to the molecule deposition chamber is a quadrupole mass spectrometer (QMS) which is used to check the cleanliness of the water vapor before deposition. A water pressure of \qty{5e-6}{\milli\bar} is introduced into the molecule deposition chamber before opening the gate valve to the preparation chamber. The surface is positioned in front of the gate valve to the molecule deposition chamber but is rotated away during opening and closing of the gate valve. Once the gate valve is open, the surface is quickly rotated to face the molecule deposition chamber for \qty{\sim10}{s} of direct deposition.
The other setup is equipped with an Oxygen Atom Beam Source (OBS, MBE Komponenten GmbH) operating at \qty{1560}{\celsius} allowing
functionalization of the surface with oxygen atoms (O atoms). It is possible to deposit water onto the surface of interest via backfilling of the preparation chamber. The surface is kept at either LN$_2$ temperature (\qty{78}{K}) or at an estimated temperature of \qty{\sim60}{K}, obtained by quickly transferring the sample between the LT-STM, operating at \qty{\sim4}{K}, and the LN$_2$ cooled manipulator. This ensures that the water is deposited before the sample has been fully reached the manipulator equilibrium temperature, followed by a quick transfer back into the STM.

Prior to each experiment, a highly oriented pyrolytic graphite (HOPG) sample was cleaved in air with adhesive tape and subsequently cleaned in UHV by heating cycles up to \qty{1000}{\kelvin}. Oxygen functionalized HOPG (O-HOPG) is produced through short \qty{\sim2}{s} exposures to O atoms from the OBS source. Deionized water was purified via several freeze-pump-thaw cycles and a mass spectrometer was used to check the water vapor for contaminations. 

All images were analyzed using Gwyddion SPM analysis software. Images presented are flattened and calibrated if needed, but all heights and distances measured are also confirmed from the unprocessed images, presented in the supporting information, to ensure that no artificial errors are introduced from processing the images.

\subsection*{Theoretical methods}
For all computational calculations, the graphite substrate was modeled as a single graphene layer with a lattice constant of \qty{2.465}{\angstrom} in a ($6\times6$) supercell. Oxidized graphene was represented by placing a single oxygen atom on a graphene bridge site, followed by a structural optimization. Ab initio energies were computed using density functional theory (DFT) with the Perdew-Burke-Ernzerhof (PBE) GGA functional \cite{Perdew_1996}, as implemented in the GPAW code \cite{Mortensen_2005, Enkovaara_2010}, and handled with the atomic simulation environment (ASE) package \cite{Hjorth_Larsen_2017}. To account for dispersion interactions, a D4 correction term \cite{Caldeweyher_2017, Caldeweyher_2019, Caldeweyher_2020} was added to the PBE energy predictions. Without this correction, water did not bond to the graphene surface.

Structure searches for small water clusters were carried out using the GOFEE algorithm \cite{Bisbo_2020, Bisbo_2022}, as implemented in the AGOX package \cite{Christiansen_2022}. Ten parallel runs with different initial seeds were done for each cluster size. The best structure from all runs was further optimized with a DFT force convergence criterion of \qty{0.01}{\eV\per\angstrom}. More details on the exact DFT settings used can be found in the supporting information.

To compare the stability of water clusters formed on bare and oxidized graphite, cluster formation energies are needed. Relative to individual water molecules far apart on the graphene surface, these formation energies are calculated as
\begin{align}
    E^\text{bare}_\text{form}(N) &= E^\text{bare}_{N \; \text{\ce{H2O}}} + (N-1)E^\text{bare} - N E^\text{bare}_\text{1 \ce{H2O}},
    \label{eq:energy_gain_cluster_no}\\
    E^\text{oxidized}_\text{form}(N) &= E^\text{oxidized}_{N \; \text{\ce{H2O}}} + (N-1)E^\text{oxidized} - N E^\text{oxidized}_\text{1 \ce{H2O}*}.
    \label{eq:energy_gain_cluster_with}
\end{align}

Here, $N$ is the cluster size, $E^\text{bare}_{N \; \text{\ce{H2O}}}$ and $E^\text{oxidized}_{N \; \text{\ce{H2O}}}$ are energies of the global minimum configurations on bare or oxidized graphene, $E^\text{bare}$ is the energy of the bare graphene surface, $E^\text{oxidized}$ is the energy of the oxidized graphene surface, $E^\text{bare}_\text{1 \ce{H2O}}$ is the energy of the global minimum configuration of a single water molecule on bare graphene, and $E^\text{oxidized}_\text{1 \ce{H2O}*}$ is the energy of a water molecule on the oxidized graphene surface far away from the oxygen. By calculating the formation energy using eq (\ref{eq:energy_gain_cluster_no}) and (\ref{eq:energy_gain_cluster_with}), all energy contributions except from \ce{H2O}-\ce{H2O} bonding and \ce{H2O}-\ce{O} bonding are subtracted.

\section{Results and Discussion}
\subsection*{H$_2$O on HOPG and O-HOPG}

\begin{figure}[t!]
    \includegraphics[width=0.95\textwidth]{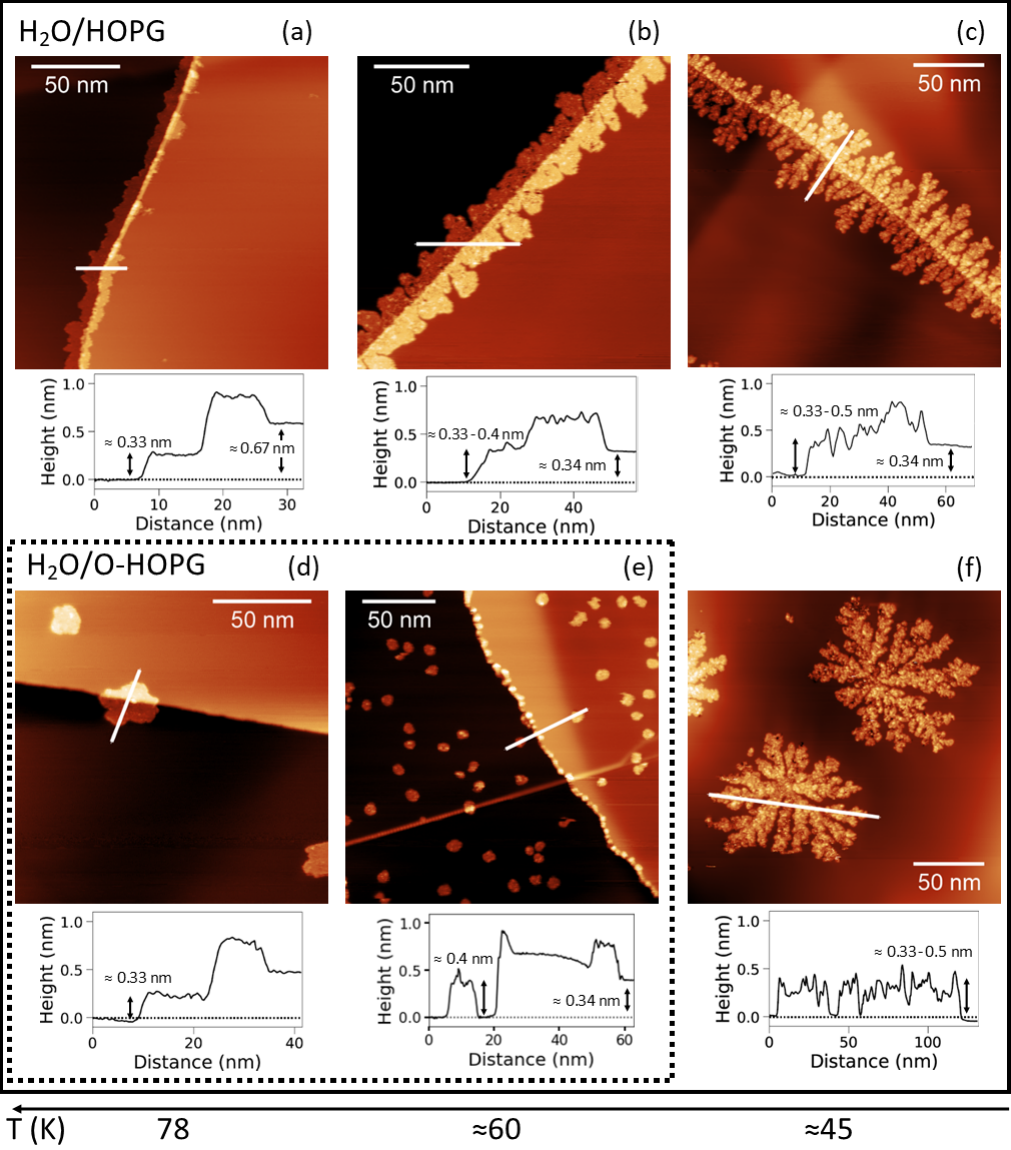}
    \centering
    \caption{Water ice growth on HOPG vs. O-HOPG at different surface temperatures:
    (a,b) HOPG and (d,e) O-HOPG after water deposition (\qty{\sim90}{\s}, at \qty{5e-9}\,mbar). (c,f) Water deposited onto HOPG via direct deposition from a molecular chamber kept at \qty{5e-6}\,mbar for \qty{\sim10}{\s}. The white lines in the LT-STM images indicate the apparent height profiles depicted below the images. Surface temperature during water deposition: (a,d) \qty{78}{K}, (b, 
    e) \qty{\sim60}{\K}, (c, f) \qty{\sim45}{\K}. Scanning temperature: (a,d) \qty{78}{K}, (b, 
    e) \qty{4}{K}, (c, f) \qty{8}{K}. Scanning parameters: I=\qty{3}{\pico\ampere}--\qty{25}{\pico\ampere} and V=\qty{4.5}{\volt}. See the supporting information for original data.} 
    \label{fig:Fig1-steps}
\end{figure}

Figure \ref{fig:Fig1-steps}(a) and (b) display STM images of pristine graphite exposed to water (\qty{\sim90}{\s}, at \qty{5e-9}\,mbar), resulting in sub-monolayer coverage, at different surface temperatures of \qty{78}{\K} and \qty{\sim60}{\K}, respectively. 
In both images, a region of the graphite surface containing a step edge is depicted. While no water ice clusters are observed on the terraces, the step edge is in both cases observed to be decorated with water clusters. Height profiles along the white lines drawn in each STM image, reveal the expected height difference of \qty{\sim3.4}{\angstrom} between the two terraces on HOPG in \ref{fig:Fig1-steps}(b) and a double step in \ref{fig:Fig1-steps}(a). Along the step-edge, the deposited water ice displays features with apparent heights of \qty{\sim3.3}{\angstrom}. The water deposited on HOPG at the temperature of \qty{\sim60}{\K}, see  Figure \ref{fig:Fig1-steps}(b), appears more dendritic and the height profile is more irregular than that deposited at the higher temperature, see Figure \ref{fig:Fig1-steps}(a), where a more compact and smooth water layer is observed.

Presented in Figure \ref{fig:Fig1-steps}(c) and (f) are STM images of an HOPG surface exposed for \qty{\sim10}{\second} to H$_2$O molecules from an attached molecular chamber at \qty{5e-6}{\milli\bar} at an estimated surface temperature of \qty{\sim45}{\kelvin}. The low temperature and the high flux lead to dendritic growth. A significant fraction of the adsorbed water molecules are located along step edges on the HOPG as shown in Figure \ref{fig:Fig1-steps}(c) and in the large clusters with diameters of several hundred nanometers on the terraces, see Figure \ref{fig:Fig1-steps}(f). 

A general observation can be made across the height profiles drawn in Figure \ref{fig:Fig1-steps}(a,b,c,f). At higher growth temperatures, the water clusters are denser and display a smoother height profile. At low temperatures, the clusters are dendritic and their height profiles vary from largely \qty{\sim3.3}{\angstrom} to \qty{\sim5}{\angstrom}. The apparent height of the water layer measured by STM is known to depend on the bias voltage \cite{stahler2007impact,mehlhorn2009height} and hence does not directly represent the morphological height of the water clusters. It is therefore not straightforward to determine whether the observed water clusters are single, double, or multilayer structures based on the measured heights. However, the variation in height suggests multilayer structures at low temperatures and at least two layers for smoother structures at higher temperatures. 

The dendritic structures in \ref{fig:Fig1-steps}(c) and \ref{fig:Fig1-steps}(f) indicate diffusion limited aggregation (DLA) where water molecules diffuse on the graphite surface until they hit and stick to a water cluster with no or limited subsequent edge diffusion \cite{witten1983diffusion}. The clusters in \ref{fig:Fig1-steps}(f) yield a fractal dimension, $D$ = 1.7, as expected for systems exhibiting diffusion-limited aggregation, see more details in the supporting information. 
The dendritic, diffusion limited growth is a result of the low temperature, which limits the energy available for restructuring/diffusion when a new water molecule is added to the existing water cluster. The low surface temperature and the relatively high deposition flux both contribute to the nucleation and growth of stable water clusters on the terraces observed in Figure \ref{fig:Fig1-steps}(f). A lowering of the temperature will have an exponential effect on the residence time of small water clusters on the terraces. An increase in deposition rate would enable faster cluster growth resulting in increased binding energies via Van der Waals force interactions with the surface which scales linearly with the number of added water molecules.

The impact of surface defects in the form of oxygen atoms chemisorbed on the HOPG surface (O-HOPG) on the ice cluster nucleation and growth is investigated in Figures \ref{fig:Fig1-steps}(d) and \ref{fig:Fig1-steps}(e). Scanning on O-HOPG requires low voltage conditions as voltages above \qty{3}{\volt} tend to desorb the oxygen. Amorphous ice, however, has poor conductivity, and to avoid tip-water interaction, a high bias voltage (above \qty{4}{\volt}) and small tip current (below \qty{10}{\pico\ampere}) were used while scanning. Hence, under the scanning conditions used in Figures \ref{fig:Fig1-steps}(d) and \ref{fig:Fig1-steps}(e) no O atoms are observed (see Figure \ref{fig:Fig1-Overview}(a) for an STM image of O-HOPG and Fig S1 in supporting information for illustration of tip-water cluster interactions on O-HOPG).
Figures \ref{fig:Fig1-steps}(d) and \ref{fig:Fig1-steps}(e) display STM images of oxygen functionalized graphite (O-HOPG) exposed to \qty{5e-9}\,mbar of water for \qty{\sim90}{\s}, resulting in sub-monolayer coverage, at surface temperatures of \qty{78}{\K} and \qty{\sim60}{\K}, respectively. Figure \ref{fig:Fig1-steps}(d) depicts a bare step edge as well as water clusters on the terraces and one cluster across step edge. The clusters in Figure \ref{fig:Fig1-steps}(d), though smaller, appear denser and less dendritic than those in Figure \ref{fig:Fig1-steps}(f), which were grown at low temperatures. A single cluster is observed on the step (a double step in this case) of Figure \ref{fig:Fig1-steps}(d). Interestingly, the cluster retains its shape across the step meaning that the water either crosses the step or interacts between plateaus. Further insight into the cluster growth dynamics is gained from this cluster extending across the step in \ref{fig:Fig1-steps}(d), which is most likely centered on an O atom defect attached to the step edge, when compared to the water which is relatively evenly spread out along the step in \ref{fig:Fig1-steps}(a). Firstly, less water reaches the step edge as it can cluster around the O atom defects on the terraces and, secondly, water molecules reaching the step can diffuse along the step and reach the cluster. Hence, more water is expected to reach the cluster from the step direction and one would expect an even distribution of water across the step similar to that observed in Figures \ref{fig:Fig1-steps}(a,b,c,e). The cluster retains a circular shape, however, which again indicates that the water diffuses along the water cluster edge or that the water molecules, upon incorporation into the cluster, have enough energy to restructure the cluster.

Decreasing the surface deposition temperature from \qty{\sim78}{\K}, Figure \ref{fig:Fig1-steps}(d), to \qty{\sim60}{\K}, Figure \ref{fig:Fig1-steps}(e), results in the formation of much smaller water clusters spread across the terraces. The step edge is also observed to be decorated with small water clusters on the upper terrace. A defect line, seen in the HOPG horizontally across the image, does not appear to influence the water clustering to any high degree. The bright area on the upper terrace towards the top of the image is presumably the result of a subsurface step in the layers below. A more in-depth investigation of how the water clusters form around the O atoms on the surface is presented in the next section to better understand the observed distributions and growth regime.

\subsection*{H$_2$O distribution on O-HOPG}
\begin{figure}[t]
    \includegraphics[width=1\textwidth]{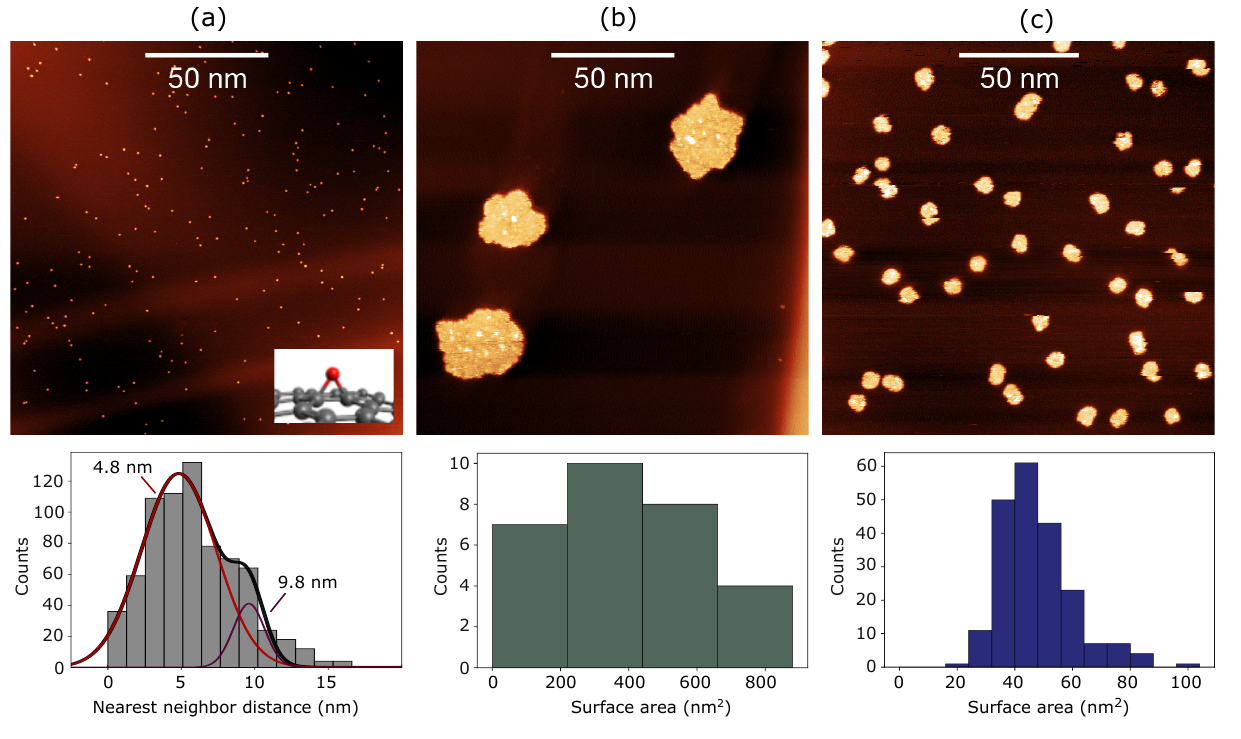}
    \centering
    \caption{Water cluster distribution on O-HOPG: (a) LT-STM image of O-HOPG. The scattered bright dots in the STM image are the O atoms adsorbed as epoxy groups on the HOPG surface, see inset scheme. A nearest neighbor analysis for the oxygen atoms, with a fit (black line) to the distribution using two normal distributions (red and violet), is presented below. (b) LT-STM image of the O-HOPG surface following deposition of water (\qty{\sim90}{\s}, at \qty{5e-9}\,mbar) at \qty{78}{\kelvin}, imaged at \qty{78}{\kelvin}. (c) LT-STM image of the O-HOPG surface following deposition of water (\qty{\sim90}{\s}, at \qty{5e-9}\,mbar) at \qty{\sim60}{\kelvin}, imaged at \qty{4}{\kelvin}. Size distributions of all imaged water clusters, including other STM images, are presented at the bottom of (b) and (c). Scanning paremeters: (a) I=\qty{1}{\nano\ampere} and V=\qty{2.26}{\volt} (b,c) I=\qty{3}{\pico\ampere}--\qty{5}{\pico\ampere} and V=\qty{4.5}{\volt}.}
    \label{fig:Fig1-Overview}
\end{figure}

Figure \ref{fig:Fig1-Overview} presents STM images exploring how the early stages of water ice growth proceeds on O-HOPG at different temperatures. 
Figure \ref{fig:Fig1-Overview}(a) displays an STM image of O-HOPG with an O atom coverage of $\sim$\qty{0.04}{\percent} (full coverage defined as one O atom per graphite unit cell). Atomic oxygen chemisorbs on HOPG as epoxy groups \cite{Larciprete2012}, see inset of \ref{fig:Fig1-Overview}(a). A nearest neighbor analysis shows that the O atoms are uniformly distributed across the surface with an average distance of \qty{\sim5}{\nano\meter}, see Figure \ref{fig:Fig1-Overview}(a). The distribution, interestingly, yields two distinct maxima, one at a distance of \qty{\sim4.8}{\nano\meter} and one at approximately double the distance of \qty{\sim9.8}{\nano\meter}, both highlighted by the fitted normal distributions. Such a double peak distribution for the nearest neighbor distance has been seen for other systems involving oxygen adsorbed onto surfaces, e.g. Ag(100) \cite{schintke2001far,sprodowski2010dissociation}, but has yet to be fully understood theoretically.
The bin size, $w$, of each histogram was optimized via the formula $w=3.49\sigma N^{-1/3}$, where $N$ is the number of data points and $\sigma$ is their standard deviation, following \cite{scott1979optimal}.

Figure \ref{fig:Fig1-Overview}(b) and (c) show STM images of O-HOPG exposed to water (\qty{\sim90}{\s}, at \qty{5e-9}\,mbar) as in Figures \ref{fig:Fig1-steps}(d) and (e), respectively. On the oxygen-functionalized HOPG, water ice clusters form on the terraces. This finding contrasts the observations for the pristine HOPG surface at similar temperatures (see Figures \ref{fig:Fig1-steps}(d) and (e)), where water clusters only at step edges. This difference indicates that O atoms limit the water mobility on the surface, acting as nucleation centers for ice growth.

Size distributions of the water clusters for both surface temperatures are presented alongside the STM images in Figure \ref{fig:Fig1-Overview}(b) and (c). The sizes of the water clusters depend greatly on adsorption temperature. For ice clusters grown at \qty{78}{\kelvin}, seen in \ref{fig:Fig1-Overview}(b), the size distribution range from \qty{100}{\nano\metre\squared}--\qty{800}{\nano\metre\squared}, with an average of \qty[separate-uncertainty = true]{400(193)}{\nano\metre\squared}. Water deposition at the lower surface temperature of \qty{\sim60}{\kelvin} (\ref{fig:Fig1-Overview}(c)) yields a much smaller and narrower size distribution. The cluster sizes are normally distributed from \qty{20}{\nano\metre\squared}--\qty{100}{\nano\metre\squared} with an average size of \qty[separate-uncertainty = true]{48(12)}{\nano\metre\squared}. These findings indicate that the water mobility on O-HOPG increases with temperature, leading to the formation of larger clusters at deposition temperatures of \qty{78}{\kelvin} than at \qty{\sim60}{\kelvin}. 

The O-HOPG surface in Figure \ref{fig:Fig1-Overview}(a) reveals a uniform distribution and an average nearest neighbor distance of \qty{\sim5}{\nano\metre}. For water deposited at \qty{60}{\kelvin}, the typical ice cluster size is \qty{\sim 48}{\nano\metre\squared}, and might therefore cover one to three oxygen atoms on average. By calculating the number density of the ice clusters, we estimate that roughly half of the O atoms are not covered by ice clusters. This indicates that a single O-H$_2$O bond is not sufficient to stabilize a water molecule at these temperatures, though it may provide a transient bond, and that water clusters are only stabilized on the terraces when they reach a sufficient size to cover several O-atoms each providing stabilizing O-H$_2$O bonds.

\subsection*{Nucleation analysis}
A capture-zone scaling (CZS) analysis \cite{pimpinelli2007capture} has been applied to better quantify the growth of clusters on the O-HOPG. It is used to determine whether the water clustering follows typical nucleation theory and provides constraints on a critical cluster size, $i^*$, where $i^*+1$ is the minimum number of molecules in a stable cluster. CZS has been successfully applied to systems with non-dendritic islands \cite{Lorbek2011}, and is therefore appropriate for this work. The O-HOPG system presented here is heterogeneous, and CZS nucleation theory is valid for homogeneous systems such as the pristine HOPG surface. However, because the individual H$_2$O molecules only bind transiently to the O atoms, we argue that the main effect of the O atoms is to decrease the diffusion rate of water clusters and molecules on the terraces, such that water-cluster formation is still governed by nucleation theory.

\begin{figure}[b!]
    \includegraphics[width=1\textwidth]{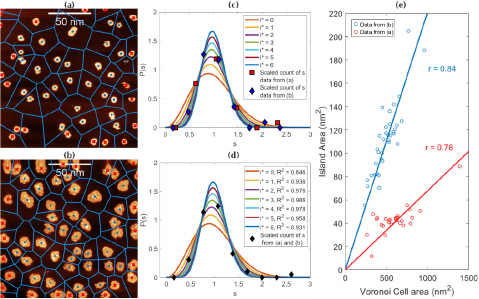}
    \centering
    \caption{CZS analysis: LT-STM images of O-HOPG after exposure to water (a) \qty{\sim90}{\s}, at \qty{5e-9}\,mbar and (b) (\qty{\sim270}{\s}, at \qty{5e-9}\,mbar), both with a surface temperature of \qty{\sim60}{\kelvin}, I=\qty{4}{\pico\ampere} and V=\qty{4.5}{\volt}. Water clusters appear as bright features with Voronoi tessellation as blue lines and the centers of each cluster is highlighted as blue dots. Red crosses indicate not-considered clusters. (c) Capture zone data for images (a) and (b), plotted against distribution functions (lines) for different critical cluster sizes, $i^*$, see equation \ref{eq:gws}. (d) Combined capture zone data for (a) and (b) presented with R$^2$ values showing a better fit for the distribution function with a critical cluster size $i^*=3$. (e) The correlation, r, between the island sizes and their Voronoi cell areas for the two different water exposures in (a), red, and (b), blue.} 
    \label{fig:VoronoiAnalysis}
\end{figure}

Presented in Figure \ref{fig:VoronoiAnalysis} is a CZS analysis of ice clusters at two different water coverages on O-HOPG. Figure \ref{fig:VoronoiAnalysis}(a) is an STM image of O-HOPG after exposure to water for \qty{\sim90}{\s}, at \qty{5e-9}\,mbar with a surface temperature of \qty{\sim60}{\kelvin}, and Figure \ref{fig:VoronoiAnalysis}(b) is an STM image of that same surface subsequently exposed to water for \qty{\sim180}{\s} (\qty{\sim270}{\s} in total), at \qty{5e-9}\,mbar with a surface temperature of \qty{\sim60}{\kelvin}.
The capture zones are determined from Voronoi tessellation around the center of each water island, see blue lines and blue crosses in Figures \ref{fig:VoronoiAnalysis}(a) and \ref{fig:VoronoiAnalysis}(b). 
Red crosses indicate clusters that are not included in the analysis as their capture zones extends beyond the STM image and cannot be calculated. Normalizing the capture-zone areas, $A$, to the average area, $<A>$, yields $\text{s}=A/\langle A\rangle$. The capture-zone histograms of the normalized areas, s, for STM images \ref{fig:VoronoiAnalysis}(a) and \ref{fig:VoronoiAnalysis}(b) are plotted in Figure \ref{fig:VoronoiAnalysis}(c) and their combined histogram in Figure \ref{fig:VoronoiAnalysis}(d). The bin size, $w$, of each histogram was optimized via the formula $w=3.49\sigma N^{-1/3}$.
Superimposed to both distributions are several theoretical distribution functions $P(s)$, described by the generalized Wigner surmise (GWS),

\begin{align}
    P(s) = a_\beta s^\beta e^{-b_\beta s^2}
    \label{eq:gws}
\end{align}
\noindent\begin{tabularx}{\textwidth}{@{}XXX@{}}
  \begin{equation}\small
  a_\beta = 2\left[\Gamma\left(\frac{\beta + 2}{2}\right)\right]^{\beta+1}\biggr/\left[\Gamma\left(\frac{\beta+1}{2}\right)\right]^{\beta+2}
    \label{eq:gws a}
  \end{equation} &
  \begin{equation}
  b_\beta = \left[\Gamma\left(\frac{\beta + 2}{2}\right)\biggr/\Gamma\left(\frac{\beta+1}{2}\right)\right]^{2}
    \label{eq:gws b}
  \end{equation}
\end{tabularx}
in which the only parameter $\beta = i^* + 2$, for a two-dimensional system, governs the variation in shape \cite{shi2009capture,pimpinelli2010pimpinelli,li2010comment}. The best fit to the experimental data in Figure \ref{fig:VoronoiAnalysis}(d) is found through an R$^2$ analysis which yields a better fit for $i^*=3$. Hence, a cluster nucleus with a size of at least four H$_2$O molecules is needed to start island formation according to the CZS analysis. 

The validity of the capture zone model can be determined using the correlation coefficient, $r$, between the individual island area and its capture zone determined from the Voronoi cell area. Figure \ref{fig:VoronoiAnalysis}(e) plots the surface area of a water cluster against its Voronoi cell area for the two different water coverages. The data points are fitted with a linear regression to guide the eye and help find $r$. With increasing coverage, $r$ approaches unity which is expected in the capture zone model as the area of each cluster approaches its Voronoi cell area. Hence, the aggregation of incoming water molecules appears to follow typical capture zone dynamics, meaning that water molecules tend to incorporate into the nearest cluster on the surface. 

The water ice clusters, which have been studied for several hours to days in the STM at \qty{\sim4}{\kelvin} and underwent temperature increases to \qty{\sim60}{\kelvin} during subsequent water depositions, as seen for Figure \ref{fig:VoronoiAnalysis}(a) to \ref{fig:VoronoiAnalysis}(b), show no indications of Ostwald ripening. Hence, the water-water interaction is high and the water molecules stick to and stay with the first available cluster they meet on the surface. 

\subsection*{Cluster energetics and dynamics}

\begin{figure}[t]
    \centering
    \includegraphics[width=1\textwidth]{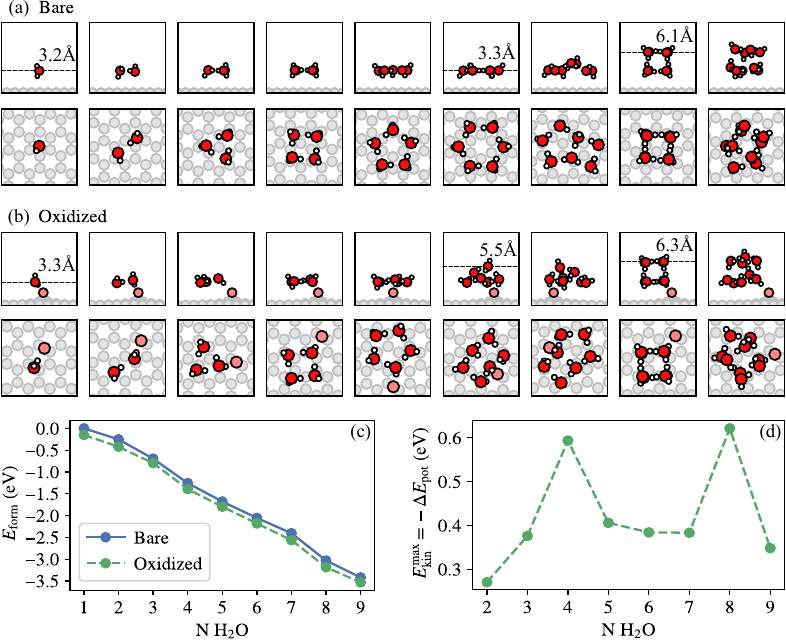}
    \caption{Optimal water clusters found using GOFEE on (a) bare graphene and (b) oxidized graphene. (c): Formation energies of the clusters calculated using eq. (\ref{eq:energy_gain_cluster_no}) and (\ref{eq:energy_gain_cluster_with}). (d): For oxidized graphene: Potential energy gained from the addition of a single water molecule to a cluster as a function of the resulting cluster size.}
    \label{fig:clusters_and_energy_gain}
\end{figure}

To better understand the energetics of water cluster nucleation on the oxidized graphite, we identified the global minimum water configurations on bare and oxidized graphene using the machine learning-accelerated structure search algorithm GOFEE. Searches were performed with one to nine water molecules, and the resulting structures are shown in Figure \ref{fig:clusters_and_energy_gain}(a) and (b) for bare and oxidized graphene, respectively. The oxygen on the graphene surface is shown in a lighter shade compared to oxygen belonging to water, for easier distinguishability. The most stable water clusters found are seen to contain four-, five-, or six-membered motifs of water molecules exhibiting hydrogen bonding between neighboring molecules. Clusters consisting of seven or more water molecules contain multiple such motifs. No seven-membered motifs were observed in any of the structure searches with seven or more water molecules, indicating that geometries consisting of multiple motifs are more stable than a single large ring at larger cluster sizes. Similar structures are known to act as precursors of ice nucleation at the surface of homogeneous systems, such as water nanodroplets \cite{Sun_2024}.  For all cluster sizes, except for the six-water cluster, the cluster geometries are similar on bare and oxidized graphene. The difference in all cluster structures is the added hydrogen bonds between the water molecules and the oxygen on the surface. For cluster sizes up to 5 H$_2$O molecules single layer clusters with heights of \qty{\sim3.3}{\angstrom} are formed, while for larger clusters increased heights are found until at 8 H$_2$O molecules a double layer with a height of \qty{\sim6.1}-\qty{6.3}{\angstrom} is formed. 

Insight into the stabilizing effect of oxygen atoms was obtained by comparing cluster formation energies, calculated using eq. (\ref{eq:energy_gain_cluster_no}) and (\ref{eq:energy_gain_cluster_with}), for the identified clusters. Formation energies for all clusters are presented in Figure \ref{fig:clusters_and_energy_gain}(c). The energy difference between cluster formation on bare or oxidized graphene ranges from \qtyrange{0.10}{0.17}{\eV}, which is comparable to the energy of a hydrogen bond. The formation energy differences are also presented in Table S1 of the supplementary information. Inspection of all clusters on oxidized graphene reveals the presence of such a hydrogen bond between the water clusters and the oxygen. The hydrogen bonds thus makes it energetically favorable for small water clusters to form near oxygen atoms on the surface compared to bare graphite. The potential energy change upon addition of a single water molecule to a cluster was calculated as
\begin{equation}
    \Delta E_{\text{pot}}(N) = E_{\text{form}}(N) - E_{\text{form}}(N-1).
\end{equation}
The change in potential energy for all cluster sizes, equivalent to the maximum kinetic energy the cluster can gain from the water addition, is presented in Figure \ref{fig:clusters_and_energy_gain}(d) for the oxidized graphene. The energy varies from \qtyrange{0.3}{0.4}{\eV} for most cluster sizes and reaches \qty{\sim0.6}{\eV} when a four- or eight-membered cluster is grown, showing that the stability of these clusters per water molecule is uniquely high.
The large gain in potential energy when going from \num{3} to \num{4} water molecules is also in agreement with the critical cluster size $i^*=3$ found from the CZS analysis in the experimental section. 

\begin{figure}[t]
    \centering
    \includegraphics[width=1\textwidth]{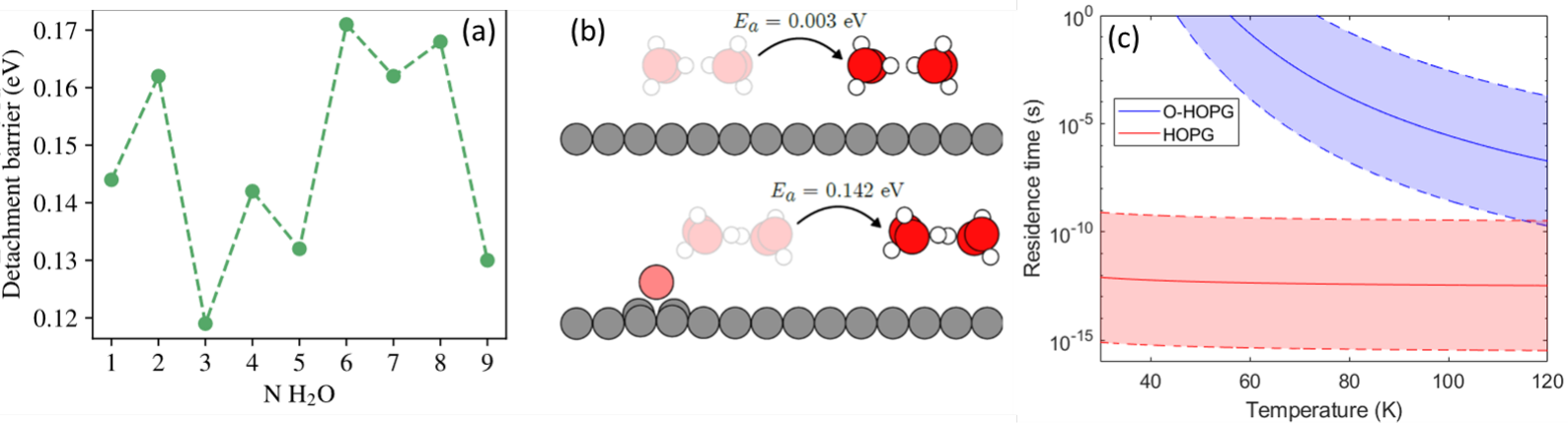}
    \caption{(a): Detachment barrier sizes for different cluster sizes calculated using the nudged elastic band method (b): Barrier sizes calculated for a 4-membered ring to diffuse on clean graphene (upper panel), and detachment from oxygen (lower panel). (c) The calculated residence time, see equation (\ref{eq:residence time}), for a 4-membered water cluster when attached to the pristine HOPG surface and to an O atom on the O-HOPG surface.}
    \label{fig:energy_released_and_barriers}
\end{figure}

A deeper understanding of cluster dynamics on the surface was achieved by computing barriers for cluster diffusion and detachment from oxygen defects using the nudged elastic band (NEB) method \cite{Kolsbjerg_2016}. Diffusion barriers for clusters on bare graphene were found to range from \qtyrange{0.001}{0.03}{\eV}. These small barriers are in line with previous computational studies of \ce{H2O} diffusion on graphene \cite{Ma_2011, Ma_2015}, and also match the high mobility of singular or small clusters of water molecules observed in the experiments. The detachment barriers for clusters bonded to oxygen defects, presented in Figure \ref{fig:energy_released_and_barriers}(a), range from \qtyrange{0.12}{0.17}{\eV}.

If the potential energy change, $\Delta E_{\text{pot}}$, upon adding an water molecule to an already formed water cluster is converted into kinetic energy of the cluster (before eventually being lost to phonons of the graphene), a total of \qtyrange{0.3}{0.6}{\eV} is available. The maximal kinetic energy that the cluster can gain, $E_{\text{kin}}^{\text{max}}$, also presented in Figure \ref{fig:clusters_and_energy_gain}(d), is seen to be sufficient for the cluster to overcome the detachment barrier and start diffusing across the surface. The small diffusion barriers further suggest that larger clusters will stay mobile on the surface, and only become fully immobilized when they attach to multiple oxygens. Assuming an attachment strength of \qty{\sim0.15}{\eV} per oxygen atom, clusters would need to bind to multiple oxygen before they are not able to overcome the detachment barrier from the energy release following the addition of a water molecule. We therefore speculate that a cluster must bind to at least two oxygen atoms before it becomes fully immobilized which could explain why roughly half the oxygen were estimated to not be covered by nanometer-sized water clusters in the experiments. The activation energy for water addition is not expected to inhibit cluster growth. Computing barriers for water addition using the NEB method showed that addition to a cluster of 3 water molecules was barrierless, while addition to a cluster of 5 water molecules was \qty{\sim0.09}{\eV}, suggesting that the energy required for water addition in general is small enough to not be of importance.

\subsection*{Residence time}\label{sec:resitime}
Given the considerations in the previous sections, it is expected that water molecules and small clusters of molecules exhibit high mobility on HOPG and cluster around larger kinks/defects such as steps on the HOPG surface. Oxygen defects on the surface are observed to stabilize water cluster nuclei on the terraces of the HOPG and initiate the formation of larger islands. 
It is not large insurmountable detachment barrier for individual H$_2$O molecules bonded to single O atoms that is responsible for the formation. Instead, it is likely an increased residence time on the terrace. The increased residence time is related to the decreased mobility of H$_2$O molecules and small clusters through transient bonds with the chemisorbed O atoms. It results in the build up of larger water clusters that will eventually overlap with several O atoms on the surface.
This in turn stabilizes the cluster as the gained energy from new incoming water molecules attaching to the cluster is not enough to detach the larger clusters from the O atoms. Presented in Figure \ref{fig:energy_released_and_barriers}(c) are simple calculations of the residence time of a water cluster on the clean HOPG and O-HOPG surface, using the diffusion barriers presented in Figure \ref{fig:energy_released_and_barriers}(b). The residence time $t_a$ is estimated via the detachment rate equation:
\begin{equation} \label{eq:residence time}
k_a = \frac{1}{t_a} = \kappa_a e^{-E_a/k_BT}
\end{equation}
for a given cluster of water molecules with binding energy $E_a$ to, e.g., an O atom on the surface at temperature $T$. The pre-exponential factor $\kappa_a$ has been sparsely studied for water diffusion on graphite, but water diffusion on graphene on Ni(111) yielded a pre-exponential factor of \qty{4e12}{\per\second}\cite{sacchi2023water}. Using $\kappa_a=$ \qty{4e12}{\per\second} yields the curves presented in Figure \ref{fig:energy_released_and_barriers}(c) with a large uncertainty range, shown as shaded areas, for the pre-exponential factor ($\kappa_a=$ \qty{4e9}{\per\second} -- $\kappa_a=$ \qty{4e15}{\per\second}). The large uncertainty range reflects predictions from other works with water diffusion and general assumptions when it comes to molecular diffusion on surfaces \cite{chen2024astrochemical,bertram2019anomalously,heidorn2015consecutive}.
The residence time goes from picoseconds on the clean HOPG to the order of microseconds on the O-HOPG at LN$_2$ temperatures. It explains why only few large water islands are observed on the terraces of O-HOPG and none on the HOPG at \qty{78}{\kelvin}. At the lower temperature of \qty{\sim60}{\kelvin}, however, the residence time increases by several orders of magnitude and the water clusters have residence times in the order of seconds next to O atoms on the HOPG surface. These long residence times on the terraces of the HOPG likely allow the clusters to grow large enough to be stabilized by several O atoms. 
A large energy gain is observed when a water molecule attaches to a cluster, see previous section in relation to Figure \ref{fig:clusters_and_energy_gain}(d), and the provided energy from adding more water molecules does not overcome the detachment barrier for the removal of an individual molecule from the cluster. Therefore, clusters likely form, stay together, and diffuse around between singular O atoms on the surface until they eventually grow large enough to stabilize around several O atom defects on the surface. 

At lower temperatures, as those observed in the ISM reaching \qty{\sim10}{\kelvin}, the effect will be enhanced dramatically and even smaller clusters may freeze out around defects on grain surfaces. This distributes the water more sparsely across the surface and increases the cluster edge to cluster interior ratio.  Water is often seen as both an assistant in reactions \cite{molpeceres2021carbon} and also a possible hindrance when it comes to molecular diffusion/rotation inside ice cavities thus introducing small barriers for otherwise barrier-less reaction pathways in gas phase \cite{enrique2019reactivity}. Hence, edge diffusion along small water clusters may be a relevant factor in activating chemistry during ice formation, as the icy dust surface provides a meeting place for radicals and can act as a third body absorbing the potential exothermic energy released in a chemical reaction. In addition, the more even distribution of water molecules may enhance chemistry involving water as a reactant, and the carbonaceous surface, increases, e.g., the formation of CO and CO$_2$ via energetic processing \cite{mennella2004formation,shi2015vacuum}.

\section{Conclusion}
The results highlight how water molecules cluster on the surface of HOPG and O-HOPG. At low temperatures \qty{\sim 45}{\kelvin} water forms dendritic cluster structures on HOPG as expected in systems governed by diffusion limited aggregation. At higher temperatures around \qty{78}{\kelvin} water molecules form less porous structures clustered around step edges on the HOPG surface. The addition of chemisorbed oxygen atoms on the HOPG (O-HOPG) surface results in reduced diffusion of water molecules and nucleation of water clusters on the terraces. A critical cluster size for water nucleation on the O-HOPG of $i^*=3$ means that at least 4 molecules are needed to form a stable cluster, in accord with the presented theoretical calculations which revealed a high energy gain for the addition of a single water molecule forming a four-membered cluster.  
The diffusion barrier of water cluster is of the order of meV, \textit{e.g.} \qty{3}{\meV} for the case of the four-membered ring, on pristine HOPG. 
The O-HOPG surface prevents the water clusters from quickly diffusing to steps or into larger clusters. However, single defects are not enough to completely stabilize smaller clusters as, \textit{e.g.}, a detachment barrier for the four-membered cluster is \qty{0.142}{\eV} which is significantly below the attachment energy of \qtyrange{0.3}{0.6}{\eV} when additional molecules attach to the cluster. The clusters are themselves strongly bound and no Ostwald ripening is not observed.

As mentioned in the introduction, Kyrkjebø \textit{et al}. \cite{Kyrkjebo2021} reported on a lower freezing onset for water droplets on oxidized graphene compared to graphene. This effect has been suggested to be due to an increased viscosity near the O-defects on the surface hindering the restructuring necessary for a phase change. This suggestion is corroborated by the results presented here demonstrating that diffusion of water on O-HOPG is hindered by the formation of transient bonds to the O atoms on the surface. This results in residence times of small water clusters and molecules around O atoms in the order of milliseconds to seconds, much longer than on the pristine HOPG where water diffusion has negligible barriers. The resulting reduced mobility may well lead to the macroscopic effect of a local, increased viscosity preventing a phase change. This effect increases as the temperature decreases and we speculate briefly that the findings here could impact interstellar chemistry, since cold interstellar dust grains may have water ice more sparsely distributed on the surface.

In summary, we show that on the microscopic level, O atoms on the HOPG surface transiently bind water molecules, thus reducing their overall mobility on the surface. This could explain the macroscopic anti-icing properties and change the initial build-up of ices on interstellar dust grains. 

\section{Data Availability}
Link to the experimental and theoretical data in this manuscript. Archived on ERDA: \href{https://www.erda.au.dk/archives/51b679823589e56ba10924803eef72f6/published-archive.html}{Data}

\section{Supporting Information}

Diagrams of the experimental setups (Figure S1). Description of scanning parameter issues, including example of tip-interaction with surface (Figure S2). Example of defect line and \ce{H2O} cluster formation around this defect (Figure S3). Unflattened STM images (Figure S4). Details on diffusion limited aggregation and a box counting analysis of clusters on HOPG (Figure S5). Crystalline \ce{H2O} structures on graphene found using global optimization (Figure S6). DFT settings and other computational details. Cluster formation energies (Table S1).

\section{Acknowledgments}
The work is supported by the Danish National Research Foundation through the Center of Excellence “InterCat” (Grant agreement no.:DNRF150), the European Union (EU), VILLUM FONDEN through Investigator grant, project no. 1656, and the Cluster of Excellence RESOLV (EXC 2033 - 390677874 - RESOLV)
funded by the Deutsche Forschungsgemeinschaft.

\section{Competing Interests}
The authors declare no competing interests.

\newpage
\bibliography{main.bib}
\bibliographystyle{elsarticle-num} 

\end{document}


\title{Water Nucleation via Transient Bonds to Oxygen Functionalized Graphite}

\author[1,*]{Frederik G. Doktor}

\author[1]{Niels M. Mikkelsen}

\author[1]{Signe Kyrkjebø}

\author[3]{Prashant Srivastava}

\author[1]{Richard Balog}

\author[1]{Bjørk Hammer}

\author[3]{Karina Morgenstern}

\author[1,2]{Liv Hornek\ae r}

\affil[1]{Center for Interstellar Catalysis (InterCat), Department of Physics and Astronomy, Aarhus University, Ny Munkegade 120, 8000 Aarhus C, Denmark}
\affil[2]{Interdisciplinary Nanoscience Center (iNANO), Aarhus University, Gustav Wieds Vej 14, 8000 Aarhus C, Denmark}
\affil[3]{Faculty of Chemistry and Biochemistry, Physical Chemistry I, Ruhr
University of Bochum, Universitätsstraße 150, D-44801 Bochum, Germany}
\affil[*]{Frederik\_doktor@phys.au.dk}

\maketitle

\section{Experimental setups}
\begin{figure}[h!]
    \centering
    \includegraphics[width=1\textwidth]{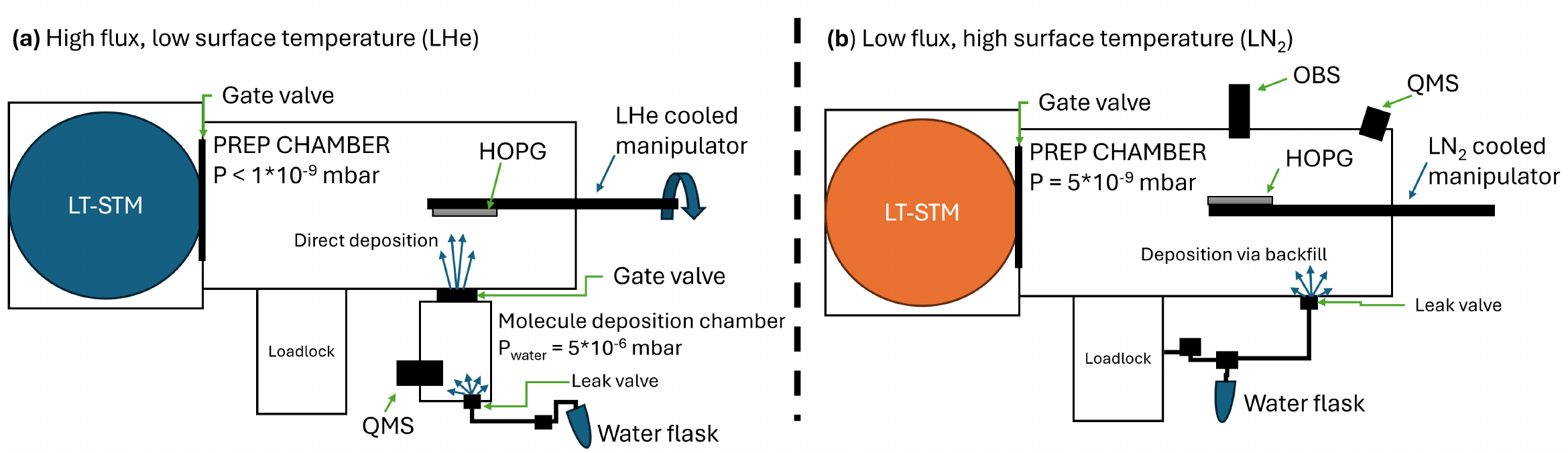}
    \caption{Sketches of the two experimental setups (birds eye view).}
    \label{fig:Appendix-Setups}
\end{figure}

Presented in Figure \ref{fig:Appendix-Setups} are two sketches of the different experimental setups utilized in this work. Figure \ref{fig:Appendix-Setups}(b) illustrates how the sample is quickly rotated to face the molecule deposition chamber for direct dosing after opening the gate valve. Figure \ref{fig:Appendix-Setups}(b) shows how the sample may face the OBS for oxidation prior to water deposition, and the water is introduced into the prep chamber for typical backfill deposition.

\section{Scanning parameter issues}
\begin{figure}[h!]
    \centering
    \includegraphics[width=0.3\textwidth]{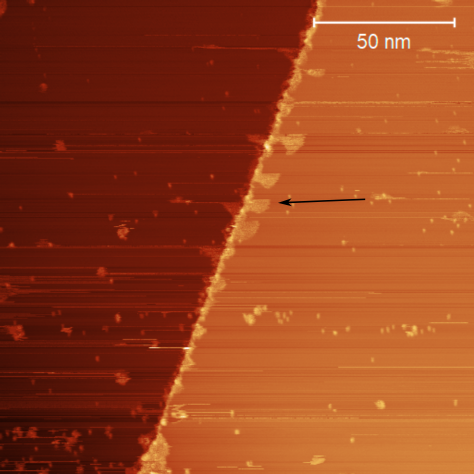}
    \caption{An STM image obtained at 4\,K of \ce{H2O} on O-HOPG scanned at V=2.9\,V and I=6.6\,pA, which allows for the O-atoms on the surface to be imaged, but results in disruptive interaction with the tip.}
    \label{fig:Appendix-Scanning}
\end{figure}

Presented in Figure \ref{fig:Appendix-Scanning} is an STM image showing how the wrong scanning parameters might disrupt the water on the surface as the tip interacts with the clusters. Tip-interaction is demonstrated as a cluster moves during the scan and/or the appearance of horizontal lines. One example of a dragged ice cluster is highlighted with the black arrow and left behind is the chemisorbed O atoms. The image was scanned at a lower bias voltage of \qty{2.9}{\volt} and clearly demonstrate interaction between the tip and the ice clusters.

\section{Pearls on a string}
\begin{figure}[h!]
    \includegraphics[width=0.6\textwidth]{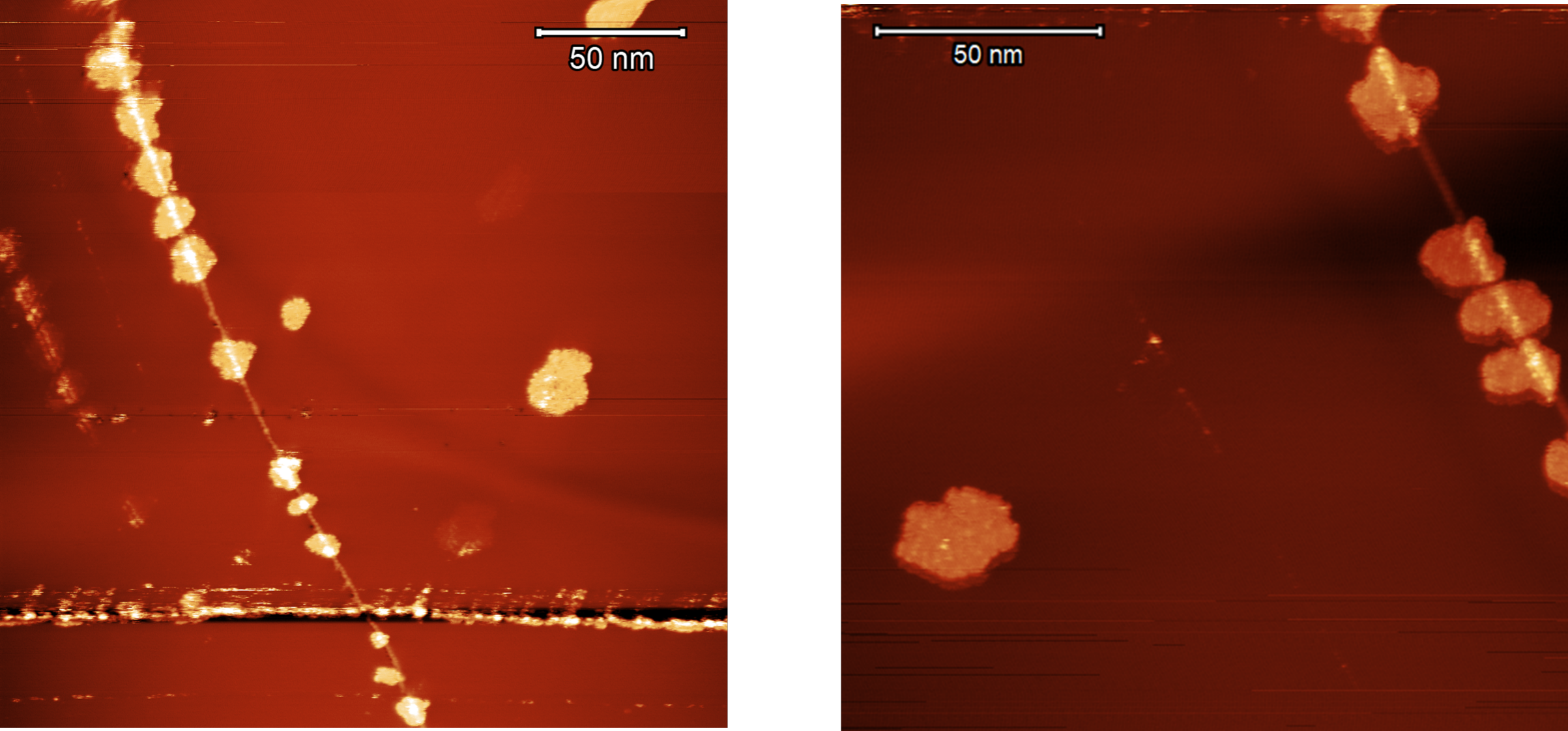}
    \centering
    \caption{STM images scanned at 4\,K of \ce{H2O} deposited at 80\,K onto O-HOPG. A defect line in the HOPG surface allows clustering. (Left) V=4.5\,V, I=4.4\,pA, (Right) V=4.5\,V, I=4.4\,pA.}
    \label{fig:Appendix-Pearls}
\end{figure}

Presented in Figure \ref{fig:Appendix-Pearls} are STM images of water deposited on O-HOPG at 80 K. The \ce{H2O} does not "wet" the defect line like on observed on the steps. Seems to indicate formation of clusters that move to the defect line and stabilize to then continue its growth. This is at higher temperature than the image with a defect line presented in the main article, hence the water clusters are more mobile and may reach this stable area on the surface. The defect line is seen as a thin bright line vertically across the image which acts as binding sites for the smaller clusters of \ce{H2O} which eventually stabilizes and forms clusters.

\section{Raw STM images}
Presented are unflattened STM images giving a more correct value of the height of certain features extracted from the height-profiles. The values ar every similar to those presented in the flattened data presented in the main text.

\begin{figure}[h!]
    \centering
    \includegraphics[width=0.8\textwidth]{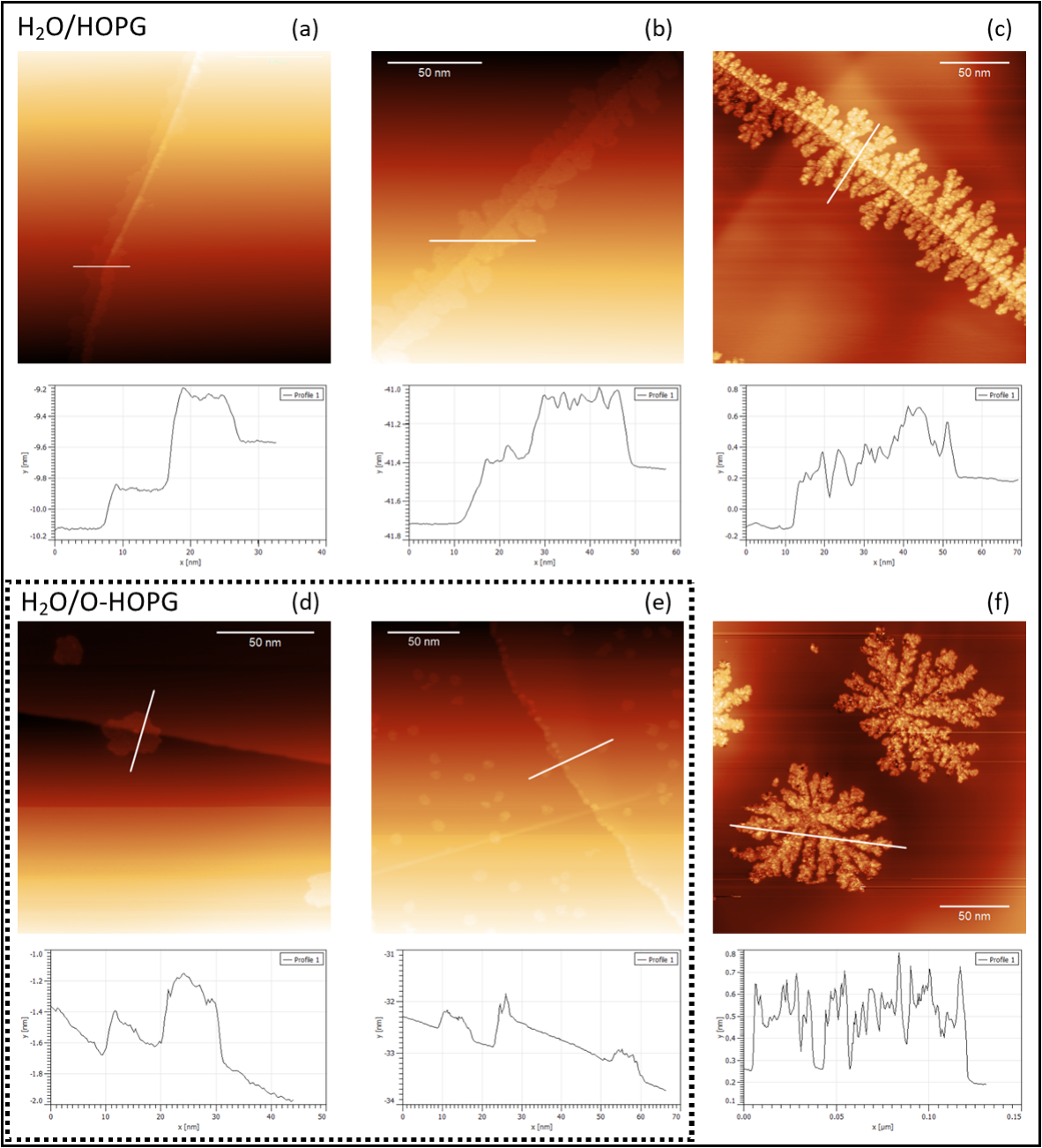}
    \caption{STM images with height scans of the different structures presented in the main text Figure 1, without any post processing. (a) V=4.5\,V, I=3.0\,pA, (b) V=4.4\,V, I=4.8\,pA, (c) V=4.5\,V, I=15\,pA, (d) V=4.5\,V, I=3.3\,pA, (e) V=4.5\,V, I=4\,pA, (f) V=4.2\,V, I=20\,pA.}
    \label{fig:Appendix-unflattened}
\end{figure}

\section{Diffusion Limited Aggregation}\label{SI-DLA}
Diffusion limited aggregation (DLA) was first described by Witten and Sander \cite{witten1981diffusion} who introduced algorithms describing particles in a 2D grid with a random walk motion. When adjacent to another particle or crystal structure the particle sticks. Definitions of the grid structure changes the outcome with square structures leading to four arms in the crystal and a hexagonal lattice leading to 6, mirroring the structure of ice crystals. Newer versions of the DLA simulations alter the shape of the resulting crystal structure, \textit{e.g.}, by introduction of edge diffusion. DLA generally produces fractal and porous structures \cite{witten1983diffusion,liu2017fractal}.

\begin{figure}[t]
    \includegraphics[width=0.4\textwidth]{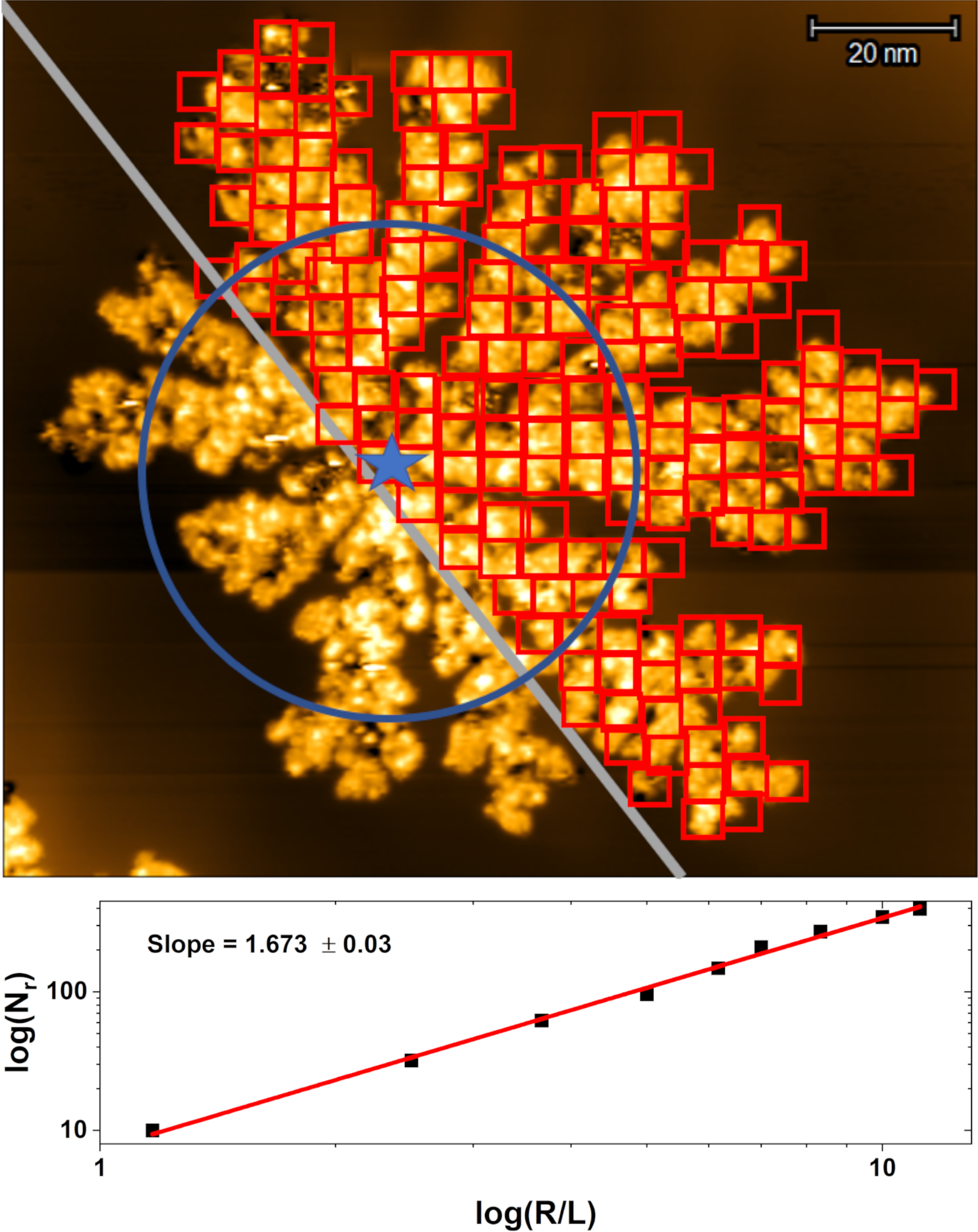}
    \centering
    \caption{A box counting analysis of a \ce{H2O} cluster on HOPG from \ref{fig:Appendix-unflattened}(f), deposited at 60\,K, to extract the fractal dimension. A neighbor cluster is the reason for the short arms in the bottom left corner which has been excluded.} 
    \label{fig:Appendix-DLA}
\end{figure}

An often-used method of DLA detection is to measure the fractal dimension, D, of the structure \cite{sander1986fractal}. The fractal dimension is an index used to characterize a change in detail of a structure relative to the change in scale. For instance, the number of particles located within a certain area may increase at a specific rate relative to the size of the area.\\
To find the number of particles within a given radius from the nucleation center, one may define a unit box n(L), with sides of length L, such that a$\ll$L$\ll$R, where a is the size of the particles, \textit{i.e.} water molecules, and R is the total radius of the island structure. The total number of unit boxes $N_r$ is then found by simply counting the boxes n(L) within a radius r.
The number of particles $N_r$ within a circle with radius r, follows a power law: 
\begin{equation} \label{eq:DLA powerlaw}
N_r \propto (R/L)^D  \quad  \Rightarrow \log(N_r) \propto D \times \log(R/L)
\end{equation}
where the slope of the log-log plot reveals the fractal dimension D. For a 2D DLA fractal structure the obtained slope value for the fractal dimension is D $\approx$ 1.7 with the exact value of D dependent on the structure of the surface grid \cite{witten1981diffusion,witten1983diffusion}. Square and triangular lattices yield a value of D = 1.66 and D=1.67, respectively, and a solid structure would have a fractal dimension approaching 2. A plot of the fitted data using equation \ref{eq:DLA powerlaw} is also shown in Figure \ref{fig:Appendix-DLA} revealing a slope of 1.67 in good agreement with a DLA fractal structure. Results were similar when using unit boxes with sides of 10 nm yielding D = 1.673 $\pm$ 0.06.

\section{Crystalline structures}
\begin{figure}
    \includegraphics[width=0.9\textwidth]{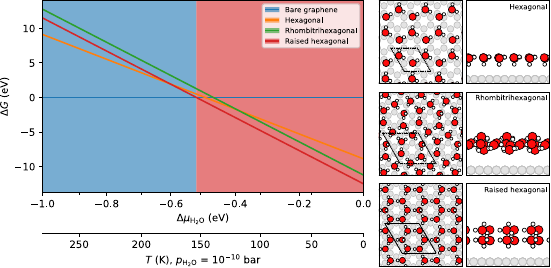}
    \centering
    \caption{Ab initio thermodynamics analysis of selected crystalline water structures on graphene. At experimental conditions, ie. low temperatures, the raised hexagonal structure found through additional structure searches is the thermodynamically most stable structure.} 
    \label{fig:aitd}
\end{figure}

To bridge the gap between cluster sizes obtained using global optimization and the observed cluster sizes in the experiment, we compared different crystalline \ce{H2O} structures on graphene from the literature \cite{Zhang2018stm, Yamada_2022}, and from a GOFEE structure search. The structure search used a ($3\times3$) supercell with six \ce{H2O}, equivalent to the configuration from \textit{Zhang et al}. Figure \ref{fig:aitd} shows all the crystalline structures considered. The hexagonal structure from \textit{Yamada et al}. consists of \ce{H2O} with their non-bonding hydrogen oriented towards the graphene surface, while the structure from \textit{Zhang et al}. forms a rhombitrihexagonal pattern. Through the GOFEE structure search, we find a structure that also forms a hexagonal pattern, but with two of the six \ce{H2O} in the hexagons slightly raised. 
Using an \textit{ab initio} thermodynamics approach, we compared the stability of the crystalline structures under experimental conditions as a function of the \ce{H2O} chemical potential. The results, presented in figure \ref{fig:aitd}, indicate that the raised hexagonal structure is the most  thermodynamically stable structure at low temperatures, whereas the other two crystalline structures are not thermodynamically stable under any conditions. Note that the list of crystalline structures considered here is not exhaustive, as more structure searches could be done using different supercell sizes and varying \ce{H2O} amounts.
The stabilizing effect of oxygen, this time for crystalline structures, was examined by performing a structure search with six \ce{H2O} and an oxygen present on a bridge site in a ($3\times3$) supercell. The global minimum structure identified was another raised hexagonal structure, with the addition of two hydrogen bonds to the surface oxygen. Using eq. (1) and (2), the formation energies were calculated to be \qty{-2.91}{\eV} and \qty{-2.82}{\eV} for the raised hexagonal structures on oxidized and bare graphene, respectively. The addition of a water molecule to the crystalline structures thus results in a energy gain of \qty{\sim0.49}{\eV} or \qty{\sim0.47}{\eV}. However, modeling the larger clusters as being purely crystalline neglects the presence of \ce{H2O} at the edge of the clusters, resulting in the actual energy gain from water addition being lower than the estimated value.

Gibbs free energies were approximated as Helmholtz free energies in the harmonic approximation at $80$ K, as implemented in the ASE thermochemistry module. The small pV term is neglected \cite{Reuter_2001}. Only the vibrational modes of the cluster were considered. Vibrational modes were evaluated by a finite difference approximation of the Hessian matrix using the high accuracy DFT settings, as implemented in the ASE vibrations module.

\section{Computational details}
All DFT calculations were done using the GPAW package \cite{Mortensen_2005, Enkovaara_2010} through the ASE package \cite{Hjorth_Larsen_2017}, with the plane-wave mode and a double zeta polarized basis set for the LCAO initialization, and the Perdew-Burke-Ernzerhof GGA functional (PBE) \cite{Perdew_1996} was used. The self-consistency cycle was stopped when the change in electron density was less than $10^{-4}$ electrons per valence electron, and the maximum change in eigenstates was less than $10^{-6}$ eV$^2$ per valence electron. Orbital occupancies were set using a Fermi-Dirac distribution with a width of \qty{0.1}{\eV}. The Brillouin-zone was sampled using $2\times2\times1$ Monkhorst-Pack sampling \cite{Monkhorst_1976}.

For the structure searches, a plane-wave cutoff energy of \qty{400}{\eV} was used. Subsequent structure optimization of the identified most stable structure from the searches and all NEB calculations were done with a plane-wave cutoff energy of \qty{500}{\eV}. These calculations also had an added force convergence criterion of \qty{0.01}{\eV\per\angstrom} for the self-consistency cycle, meaning the calculation only stopped when the difference in force for each atom was less than \qty{0.01}{\eV\per\angstrom} compared to the previous iteration.

The D4 correction term \cite{Caldeweyher_2017,Caldeweyher_2019,Caldeweyher_2020} was added to the PBE functional for all calculations, to properly include dispersion type effects in the energy calculations. Tests involving energy evaluations of an \ce{H2O} molecule at different heights above a graphene surface showed that the D4 correction was needed for an energy well to appear as function of \ce{H2O} height, meaning the PBE functional alone does not adequately describe \ce{H2O}-graphene bonding.

NEB calculations for finding diffusion barriers were done by taking the clusters found in the structure searches on bare graphene as the initial structure, and the same structure translated by the graphene lattice constant as the final structure. A similar approach was used to get the initial and final structure for calculating the barrier for detaching a cluster from a oxygen on the surface. The initial structures were the best structures found in the structure search on oxidized graphene, and the final structure was that same structure translated by a graphene lattice constant. All NEB calculations used $12$ images and the full spring force implementation described in \cite{kolsbjerg_2016}.

\section{Cluster formation energies}

Presented in Table S1 is the formation energy of all clusters found using global optimization. These are the same data used in Figure 4(c).

\begin{center}
\begin{tabular}{
    S[table-format=1]
    S[table-format=-1.2]
    S[table-format=-1.2]
    S[table-format=-1.2]
}
    \toprule
    & \multicolumn{2}{c}{$E_\mathrm{form}$ (\unit{\electronvolt})} \\
    \cmidrule(lr){2-3}
    {$N_{\ce{H2O}}$} & {Bare} & {Oxidized} & {$\Delta E_\mathrm{form}$ (\unit{\electronvolt})} \\
    \midrule
    1 &  0    & -0.15 & -0.15 \\
    2 & -0.25 & -0.42 & -0.17 \\
    3 & -0.70 & -0.80 & -0.10 \\
    4 & -1.26 & -1.39 & -0.13 \\
    5 & -1.68 & -1.80 & -0.12 \\
    6 & -2.06 & -2.18 & -0.12 \\
    7 & -2.41 & -2.57 & -0.16 \\
    8 & -3.03 & -3.19 & -0.16 \\
    9 & -3.42 & -3.53 & -0.11 \\
    \bottomrule
\end{tabular}
\captionof{table}{Formation energy data}
\end{center}

\bibliography{main.bib}
\bibliographystyle{elsarticle-num}